\documentclass[preprint,prl,aps]{revtex4}

\usepackage{graphicx}

\usepackage{amssymb}

\newcommand{\Vec}[1]{\mbox{\boldmath$#1$}}
\begin{document}




\title{A perspective of superconductivity as multiband phenomena:\\ 
Cuprate, iron and aromatic systems
}


\author{Hideo Aoki}
\address{Department of Physics, University of Tokyo, Hongo, Tokyo 113-0033, Japan\\
aoki@phys.s.u-tokyo.ac.jp}

\begin{abstract}
A theoretical overview of the classes of superconductors encompassing 
(a) high-Tc cuprate,  (b) 
iron-based and (c) aromatic superconductors is given.  
Emphasis is put on the multiband natures of all the three classes, 
where the differences in the multiorbits and their manifestations in 
the electronic structures  and pairing are clarified.  
From these, future directions and prospects are discussed. 
\end{abstract}

\maketitle


\section{Introduction}

At the occasion of the centenary of the discovery of superconductor, 
it is meaningful to give an overview of different classes of 
superconductors, which have been discovered in recent decades.  
I shall do this theoretically for the three classes of superconductors, 
namely (a) high-Tc cuprate\cite{sakakibara10},  
(b) iron-based\cite{kuroki08,kuroki09,kurokiNJP09} and (c) aromatic superconductors\cite{kosugi09,kosugi11} (Fig.1).  The purpose is two-fold: we can first look at the commonalities 
and differences among these classes of materials.   We can then 
discuss future directions and prospects for the coming decades. 

\begin{figure}[h]
\begin{center}
\includegraphics[width=11.7cm,clip]{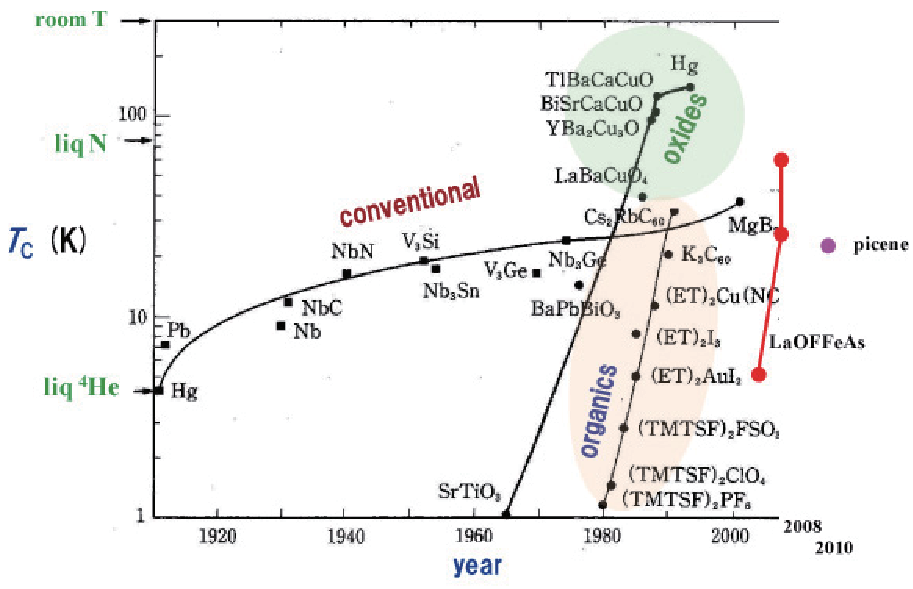}
\\
\includegraphics[width=11.7cm,clip]{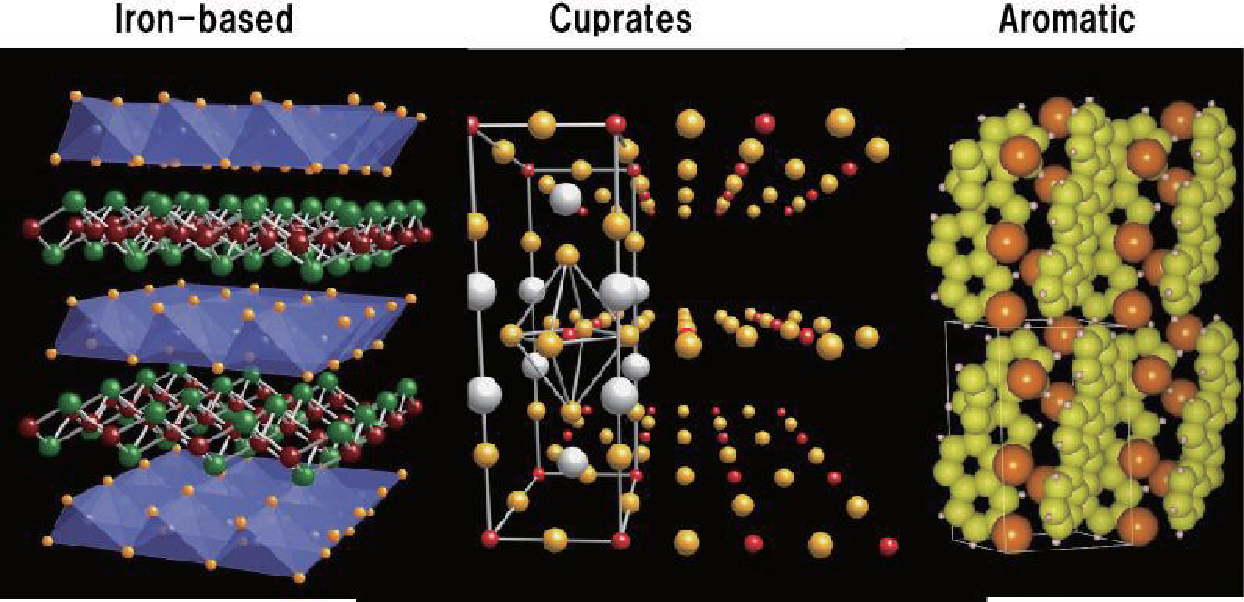}
\caption{(Top) $T_C$ plotted against calendar year 
for various classes of superconductors. 
(Bottom) Typical crystal structures of the iron-based, 
cuprate and aromatic superconductors.  
\label{fig1}}
\end{center}
\end{figure}

In the history of superconductivity, 
the conventional and intermetallic compounds were followed by 
the oxide superconductors, which culminated in the cuprates in 
the 1980's.  On the other hand, 1980 witnessed a discovery of an organic 
superconductor(SC), and 
the carbon-based SC subsequently attained 
a discovery of fullerene SC.  2008 marked the discovery of 
the iron-based SC,\cite{Hosono} the first non-copper SC with $T_C$ exceeding 50 K.  
The most recent and important 
addition to the organic SC is an aromatic superconductivity 
discovered in 2010 in K-doped picene with $T_C \simeq 20$ K.\cite{kubozono}   
So the spectrum of SC materials ranges from the 
light-element, p-electron systems (MgB$_2$, carbon-based, organic, etc) 
to transition-metal compounds (cuprates, cobaltates, 
iron-based, etc) down to higher-d-electron and f-electron systems 
(ruthenates, Hf compounds, heavy-fermion compounds, etc).  

From the viewpoint of electron mechanism of SC, 
an interest is the way in which magnetism and superconductivity appear in 
correlated electron systems, which should be sensitive to the 
factors:
\begin{enumerate}
\item underlying band structure and the associated Fermi surface,
\item spatial dimensionality, and
\item orbital degrees of freedom.
\end{enumerate}
We can even envisage a materials design by manipulating  
these factors\cite{fermiology}. 
First, the electronic structure is naturally governed by crystal structures.  
For the dimensionality, Arita et al\cite{Arita2d3d}  have shown, with 
the fluctuation exchange approximation, 
that 2D (layered) systems are generally more 
favourable for the spin-fluctuation-mediated SC 
than in 3D systems.  The reason 
is traced back to the fact that 
the spin-fluctuation-mediated pairing interaction is 
appreciable only around localised 
regions in ${\Vec k}$-space (unlike the phonon-mediated 
case), where the height and 
width of the susceptibility structures both in the frequency ($\omega$) and 
momentum (${\Vec q}$) sectors turn out to be similar between 2D and 3D.  
This means that their {\it phase volume fraction} 
is much greater in 2D, and gives higher $T_C$ in {\'E}liashberg's equation.  
This agrees with the empirical fact that almost all the 
recently discovered SC (cuprates, Co compound, iron-based, 
Hf compound, CeCoIn$_5$, solid picene, etc) are indeed layer-structured.  

A second observation is that important classes of superconductors 
such as the iron-based are often inherently multiband systems.  
We shall show that the aromatic SC also has multi-band structure 
derived from (nearly) 
degenerate molecular orbitals. So are the superconducting fullerides.  
MgB$_2$ is another multiband system, comprising $\pi$ and 
$\sigma$ bands.  Moreover, we shall show that  even the 
(single-layer) cuprates, usually regarded as a prototypical single-band 
system, can only be understood in terms of a two-band 
($dx^2-y^2, dz^2$) picture if we want to understand their 
material dependence.  Thus all the important 
superconductors are in fact multiband systems.  
Historically, multiband SC was theoretically considered 
back in the 1950's and 60's by Suhl et al and by Kondo, 
so it has a long history, but we should definitely revisit the 
physics afresh in view of the new classes of superconductors.  
In other words, multibands provide an 
important {\it internal degree of freedom} dominating the system, 
which reminds us of the hyperfine states playing an important role in cold atoms.

For each of all the three classes of materials we first construct an  
electronic model with the ``downfolding" based on 
first-principles electronic structure calculations.  
The pairing from repulsive interactions can be 
explained simply as anisotropic Cooper pairs  accompanied 
by an anisotropic gap functions, $\Delta({\Vec k})$.  
In the BCS gap equation, 
\begin{equation}
\Delta({\Vec k})=
-\sum_{{\Vec k}'}V({\Vec k},{\Vec k}')\frac{\Delta({\Vec k}')}{2E({\Vec k}')} 
{\rm tanh}\left(\frac{E({\Vec k}')}{2k_B T}\right),
\end{equation}
where $V({\Vec k},{\Vec k}')$ is the pairing interaction, 
we can immediately recognise that 
a repulsion acts as an attraction when $\Delta({\Vec k})$ 
changes sign across the typical momentum transfer 
(usually dictated by the nesting vectors that give spin fluctuation mode 
as is the case with the cuprate with an AF fluctuation 
producing a $d$-wave pairing).  
One intriguing point emerging from the relation of pairing 
with electronic structure is that 
the idea of ``disconnected Fermi 
surface"\cite{kuroki_disconnecteds} is often at work.
Namely, if the Fermi surface comprises pockets or sheets with the 
major nesting vector between the disconnected elements, we can 
realise a pairing in which {\it each pocket is fully gapped with 
a sign reversal in the gap function} between the pockets.  
This is the basic idea of the disconnected Fermi surface, 
and gives an interesting avenue for realising high Tc superconductivity.

\section{Iron-based superconductors}
\begin{figure}[h]
\begin{center}
\includegraphics[width=10.7cm,clip]{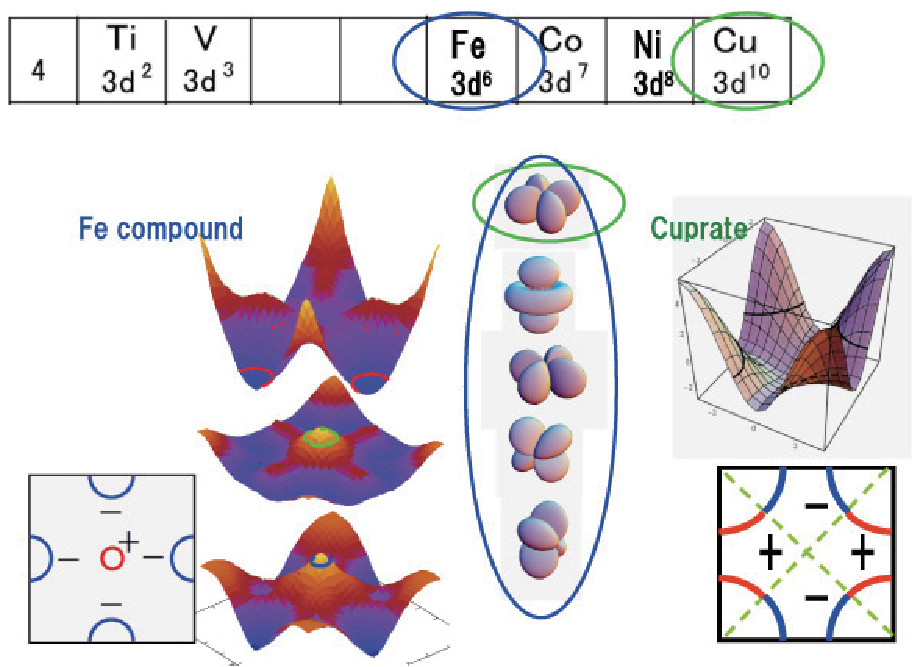}
\caption{Band dispersions and Fermi surfaces 
are schematically shown for the cuprate and 
iron-based superconductors, along with five 
d-orbitals.
\label{fig2}}
\end{center}
\end{figure}

The iron-based superconductor\cite{Hosono} is a multi-d-band system as revealed from a microscopic model construction\cite{kuroki08,mazin}.  A pairing symmetry, sign-reversing but nodeless gs$\pm$h where each Fermi pocket is fully gapped while different pockets have opposite signs, is suggested from the viewpoint of the spin-fluctuation mediated pairing.  
Thus, while an initial surprise on 
the iron-based pnictide as to why the major constituent, iron, 
which is an element usually 
associated with magnetism, we can understand that 
iron, a transition metal element situated in the middle of the periodic table, 
is characterised by multiband structures for its compounds (Fig.\ref{fig2}).    

The electronic structure calculation shows that three bands intersect $E_F$, and are 
primarily involved in the gap function, 
but actually all the five bands are {\it heavily 
entangled}, 
which reflects hybridisation 
of the five $3d$ orbitals due to the tetrahedral 
coordination of As $4p$ orbitals around Fe (Fig.1).
Hence we conclude that the minimal electronic model 
requires all the five bands. 
The Fermi surface comprises 
two concentric hole pockets ($\alpha_1$, $\alpha_2$) 
centred around $(k_x, k_y)=(0,0)$ and two electron pockets 
around $(\pi,0)$ $(\beta_1)$ and $(0,\pi)$ $(\beta_2)$. 
In addition, a portion of the band 
near $(\pm\pi,\pm\pi)$ is flat and 
close to $E_F$ around $n=6.1$, so that 
the portion acts as a ``quasi Fermi surface $(\gamma)$'' around $(\pi,\pi)$.

We then apply the five-band random-phase approximation (RPA) 
to solve the  {\'E}liashberg equation. 
We conclude that a nesting between multiple Fermi 
surfaces (pockets) results in the unconventional 
$s\pm$ pairing\cite{kuroki08}. 
Subsequently there have been a body of experimental 
results for identifying the pairing symmetry, 
and they primarily support $s\pm$, clearest of which is 
a phase-sensitive Fourier-transform STM spectroscopy\cite{hanaguri}, 
but exhibit significant 
material dependence within the iron-based compounds.

\begin{figure}[h]
\begin{center}
\includegraphics[width=12.7cm,clip]{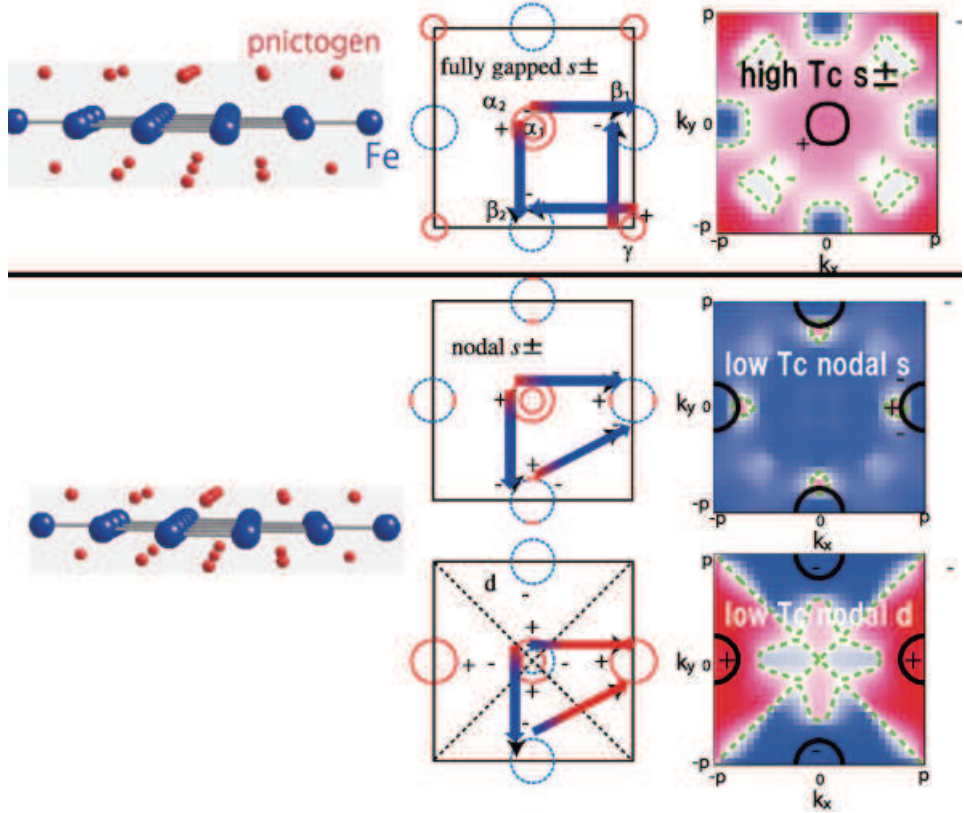}
\caption{Typical Fermi surfaces along with 
nesting vectors (middle column) and the 
gap functions (right) for higher (top) or lower (bottom)
pnictogen heights.\cite{kuroki09} 
\label{fig3}}
\end{center}
\end{figure}

Indeed, a subtlety about multiband systems appears here 
as a sensitive dependence of the pairing on materials, as revealed 
in terms of ``pnictogen height"\cite{kuroki09} (Fig.\ref{fig3}).  
The essence is that as the height is raised (e.g., as we go from 
LaFe PO to LaFeAsO or NdFeAsO), the multibands deform and 
the $X^2-Y^2$ band is shifted above $E_F$ with a new pocket emerging 
around $(k_x, k_y) = (\pi,\pi)$.  This changes the orbital character 
of the Fermi pocket and even the number of pockets.  As a result, 
LaFeAsO and NdFeAsO are happy with the $s\pm$ pairing, 
while LaFePO has multiple nesting vectors that are frustrated, 
resulting in either nodal s or d-wave pairing.  

This line of approach also explains the ``Lee plot"\cite{lee}, 
which indicates experimentally that $T_C$ attains a maximum 
when the Fe-pnictogen tetrahedron approaches a regular 
tetrahedron.  Usui and Kuroki \cite{usui} explain this 
as a maximised Fermi surface multiplicity optimising SC in iron pnictides, where 
we have a maximum number of pockets (two hole pockets and one hole 
pocket) for the  regular tetrahedron.    Impurity effects on SC should also be 
sensitive to multiband structure\cite{nagai10}.

\section{Cuprate superconductors}
Now, we can revisit the cuprate afresh, which has conventionally been viewed as a single($dx^2-y^2$)-band system.  
This comes from copper being an element situated in the right end 
of the  transition-metal periodic table.  A single-band 
structure is in fact rather rare and almost miraculous.  
Let us start from a long-standing 
puzzle --- why the single-layered cuprates, 
La$_{2-x}$(Sr/Ba)$_x$CuO$_4$ ($T_C\simeq$ 40K) 
and HgBa$_2$CuO$_{4+\delta}$ ($T_C\simeq$ 90K), have 
such a significant difference in $T_C$?  
The La system has a diamond-shaped Fermi surface 
with a relatively better nesting, while the Hg system 
has a warped Fermi surface with a poorer nesting, 
so that this sharply contradicts with a general view that 
better the nesting higher the $T_C$.  
Thus it is imperative to resolve this discrepancy 
if one wants to understand the curates.

\begin{figure}
\begin{center}
\includegraphics[width=10.0cm,clip]{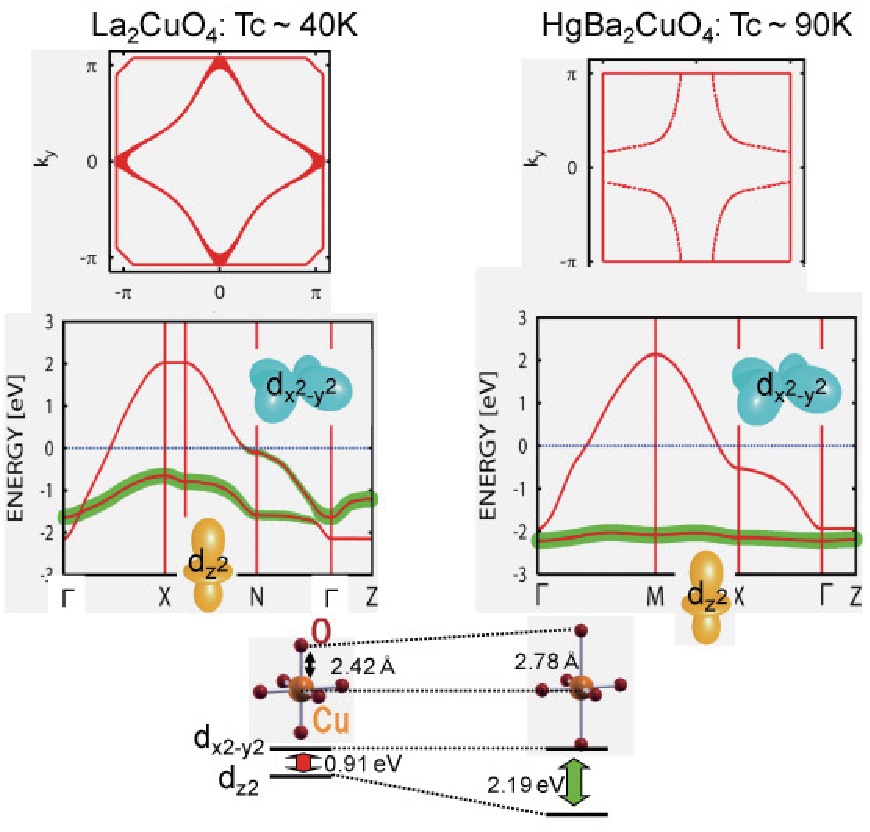}
\caption{Fermi surfaces (top) and band structures 
(middle, thickness representing the weight of $dz^2$ character) 
and the oxygen heights (bottom) are shown 
for La$_{2}$CuO$_4$ (left) and HgBa$_2$CuO$_{4}$ (right).\cite{sakakibara10} 
\label{fig4}}
\end{center}
\end{figure}

We have revealed\cite{sakakibara10}  that 
it is crucial to introduce 
a two-orbital model that explicitly incorporates the $dz^2$ orbital 
on top of the $dx^2-y^2$ (Fig.\ref{fig4}).  The former component 
has in fact a significant contribution to the Fermi surface in the 
La system, which comes from a key parameter, 
the energy level difference, $\Delta E$,  between the $dx^2-y^2$ and $dz^2$ orbitals.   A smaller $\Delta E$ 
in the La system results in simultaneously a larger 
$dz^2$ hybridisation on the Fermi surface and 
a smaller second-neighbour hopping $t_2$ (hence 
a diamond-shaped Fermi surface).   We have revealed that 
the degredation of superconductivity due to $dz^2$ hybridisation is so significant as to supersede the effect of the diamond-shaped (better-nested) Fermi surface.  This resolves the long-standing puzzle.  
Conversely, we can predict that 
materials, if any, that have smaller $t_2$ with the single-band nature preserved will realise even higher $T_C$.  

While in Ref.\cite{sakakibara10} the oxygen height above the CuO$_2$ 
plane is revealed to be a main factor, 
there exist in fact other factors.\cite{sakakibarafull}  
To be more precise, the energy difference $\Delta E$ between 
the $d_{x^2-y2}$ and $d_{z^2}$ orbitals is the key parameter that governs 
both the hybridisation and the shape of the Fermi surface.    
A smaller $\Delta E$ tends to suppress $T_c$ through a 
larger hybridisation, and vice versa.  
Then the questin is what determines the material and 
 lattice-structure dependence of $\Delta E$.   
This is shown to be 
determined by the energy difference $\Delta E_d$ between the two Cu3d orbitals (primarily governed by the apical oxygen height), 
and the energy difference $\Delta E_p$ 
between the in-plane and apical oxygens (primarily governed by the interlayer separation $d$).  

When the doping is controlled by excess or vacancy oxygen atoms, 
there is an important question of how the local change in the 
crystal symmetry caused by these oxygens affect the 
electronic structure and $T_c$, as has been 
discussed in Ref.\cite{bianconi}  
In particular, local lattice distortions will hybridise different orbitals, 
as observed by EXAFS for lattice misfits in multilayer superconductors.\cite{bianconi}  
In the present paper we have discussed how 
the material and lattice structure determine the way in which 
the orbitals are hybridised over the whole crystal, while 
the hybridisation occurs locally for local distortions.  
So a possible relation of the 
effects of apical and in-plane oxygen atoms as revealed 
here to the effects of local oxygen configurations may be an interesting 
future problem.

Incidentally, effects of electron correlation appear in a variety of ways.  
For instance, 
the evolution of the Fermi surface with pocket-like shapes as observed in 
the cuprates in ARPES experiments have theoretically been discussed in terms of the Green's function for the correlated system, in a weak-coupling scheme\cite{rice} and in the cluster extension of the dynamical mean-field theory\cite{sakai}.

\section{Aromatic superconductors}
More recently, an organic superconductivity was discovered in 
potassium-doped solid picene, which is the first aromatic superconductor 
with transition temperatures $T_C=7 - 20$ K\cite{kubozono}.  
The discovery came as a surprise, since aromatic molecules 
are most typical, textbook organic compounds, which have been certainly 
not associated with SC.  The structure is a stack of 
layers, with each layer being a herringbone arrangement 
of picene molecules.

We have obtained a first-principles electronic structure 
of solid picene as a first step towards 
the elucidation of the mechanism of the superconductivity\cite{kosugi09,kosugi11}.
The undoped crystal is found to have four conduction 
bands, which derive from LUMO (lowest 
unoccupied molecular orbital) and LUMO$+1$ 
of an isolated picene molecule, as revealed in terms of 
maximally localized Wannier orbitals.  Since there are two 
molecules per unit cell of the herringbone, we have four 
bands, which are actually entangled due to a 
hybridisation of the LUMO and (LUMO$+1$).  

\begin{figure}
\begin{center}
\includegraphics[width=12.0cm,clip]{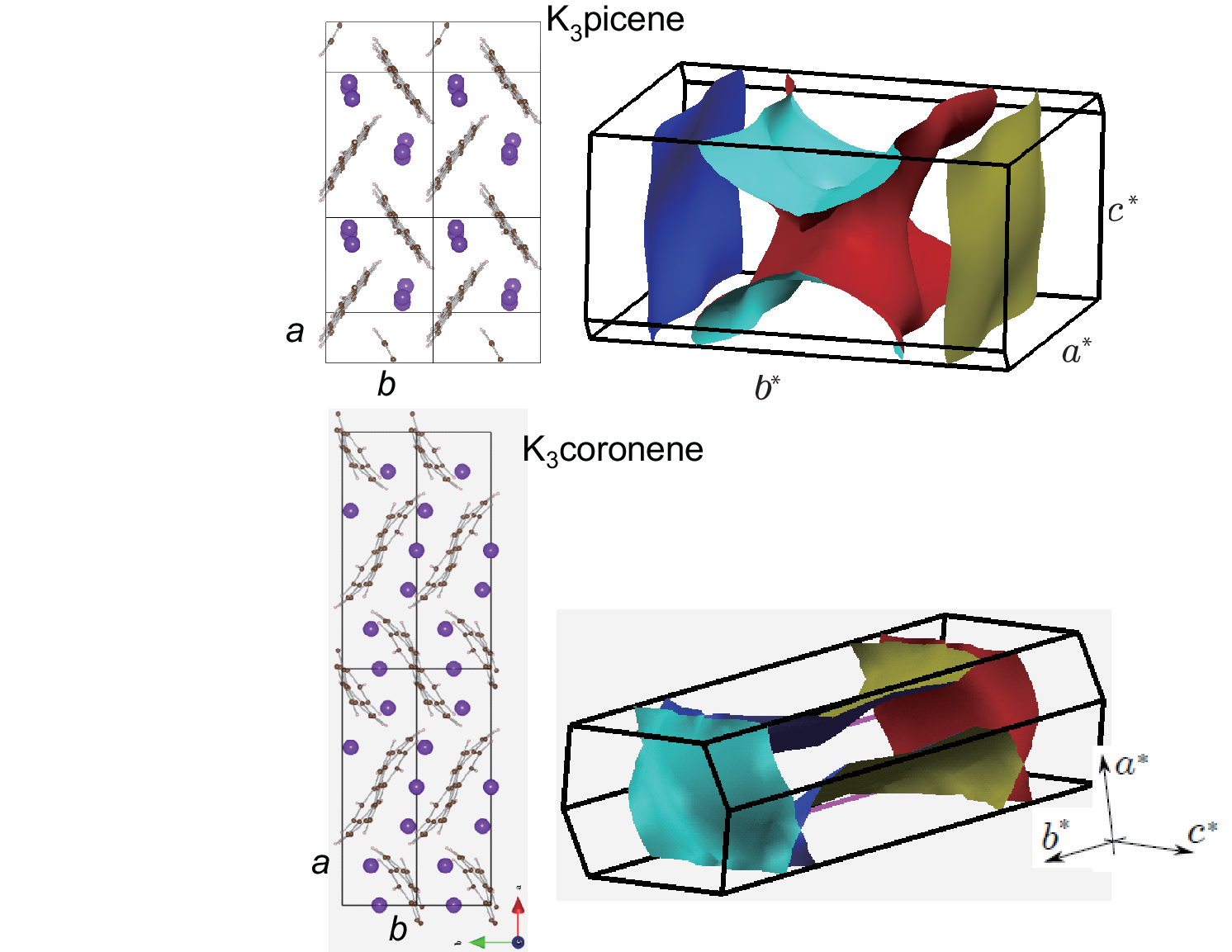}
\caption{Crystal structures (left) and Fermi surfaces (right) 
are shown for K$_3$picene (top) and K$_3$coronene (bottom).\cite{kosugi09,kosugi11}
\label{fig5}}
\end{center}
\end{figure}

For the K-doped systems, the bands are shown to be not rigid 
for two-fold reasons 
(i.e., a distorted herringbone structure upon doping along 
with a spilling of the molecular wave function over to 
potassium sites).  
The Fermi surface for K$_3$picene 
is a curious composite of a warped two-dimensional 
surface and a three-dimensional one (Fig.\ref{fig5}).

Kubozono's group has also reported superconductivity 
in another aromatic molecule, coronene, with K-doping\cite{kubozono_coronene}.  Motivated by this 
we have also obtained the first-principles electronic structure of solid coronene (Fig.\ref{fig5}).  In the undoped coronene crystal, 
the conduction band again comprises four bands, which 
originate from LUMO 
 (doubly-degenerate, reflecting a higher symmetry of the molecule) 
and entangled as in solid picene.   The Fermi surface for a candidate 
of the structure of K$_x$coronene with $x=3$, for 
which superconductivity is found,  comprises multiple sheets, 
as in doped picene but exhibiting a larger anisotropy with different topology.

As for SC, in a phonon mechanism the problem becomes the coupling 
between the electrons on such a Fermi surface and the molecular 
phonons.  The situation gives an interesting possibility for electron 
mechanisms as well, since the nesting between multiple Fermi 
surfaces can give rise to a unique opportunity for an electron 
mechanism, especially in multiband cases.\cite{fermiology}  
One might wonder whether aromatic systems can be strongly correlated, 
but a possibility of Mott transition has been discussed for K$_x$pentacene 
\cite{craciun}

\section{Discussions and summary}

Thus all of iron, cuprate and  aromatic superconductors have to be conceived as multiband systems, with some important differences among them.  
There are a host of other physical properties that are relevant to 
multiband systems.  These include (i) collective (phase) modes in multiband SC 
which become instable when three (or more) bands are frustrated\cite{ota}, and 
(ii) spin Hall effect in multiband systems\cite{pandey}.  
One important factor is the relevant energy scale in the pairing 
mechanism (e.g., the energy scale of the spin-fluctuation) 
for various classes of materials.   
 These observations should give a renewed perspective on how superconductivity can dramatically depend on elements/crystal structures.   
In a completely different avenue, we can also 
explore non-equilibrium SC\cite{tsuji11}.   
These will provide guiding principles for searching new high-Tc superconductivity for the decades to come.  

The work described here is a collaboration with 
Kazuhiko Kuroki, Seiichiro Onari, 
Ryotaro Arita, Hidetomo Usui, Yukio Tanaka, 
Hiroshi Kontani, Hirofumi Sakakibara, Taichi Kosugi, Takashi Miyake, 
Shoji Ishibashi.  
This study has been supported in part by 
Grants-in-Aid for Scientific Research from  MEXT of Japan, 
and from JST-TRIP.

%


\begin{thebibliography}{00}


\bibitem{sakakibara10} H. Sakakibara, H. Usui, 
K. Kuroki, R. Arita and H. Aoki, Phys. Rev. Lett.  
{\bf 105}, 057003 (2010).

\bibitem{kuroki08} K. Kuroki, S. Onari, R. Arita, H. Usui, Y. Tanaka, 
H. Kontani, H. Aoki, 
Phys. Rev. Lett. {\bf 101} (2008) 087004; 
{\bf 102}, 109902(E) (2009).

\bibitem{kuroki09} K. Kuroki, H. Usui, S. Onari, R. Arita and 
H. Aoki, Phys. Rev. B {\bf 79}, 224511 (2009).

\bibitem{kurokiNJP09} K. Kuroki et al, 
New J. Phys. {\bf 11}, 025017 (2009); 
H. Aoki, Physica C {\bf 469}, 890 (2009). 

\bibitem{kosugi09} T. Kosugi, T. Miyake, S. Ishibashi, 
R. Arita and H. Aoki, J. Phys. Soc. Jpn {\bf 78}, 113704 (2009); 
Phys. Rev. B {\bf 84}, 214506 (2011).

\bibitem{kosugi11} T. Kosugi et al, Phys. Rev. B (R) 
{\bf 84}, 020507(R) (2011).

\bibitem{Hosono}
Y. Kamihara, T. Watanabe, M. Hirano, H. Hosono: J. Am. Chem. Soc. {\bf 130} (2008) 3296; 
H. Takahashi, K. Igawa, K. Arii, Y. Kamihara, M. Hirano, H. Hosono: 
Nature {\bf 453} (2008) 376.

\bibitem{kubozono} 
R. Mitsuhashi et al, 
Nature {\bf 464}, 76 (2010).

\bibitem{fermiology} H. Aoki, Physica C {\bf 437-438}, 11 (2006); 
H Aoki in {\it Condensed Matter Theories} {\bf 21} 
ed. by H. Akai et al, (Nova Science, 2007).

\bibitem{Arita2d3d} R. Arita, K. Kuroki, H. Aoki: 
Phys. Rev. B {\bf 60} (1999) 14585; 
P. Monthoux, G. G. Lonzarich, Phys. Rev. B {\bf 59} (1999) 14598.

\bibitem{kuroki_disconnecteds} K. Kuroki, R. Arita: Phys. Rev. B {\bf 64} (2001) 024501; Phys. Rev. B {\bf 66} (2002) 184508.

\bibitem{mazin} 
I. I. Mazin, D. J. Singh, M. D. Johannes and M. H. Du, Phys. Rev. Lett. {\bf 101} (2008) 057003.

\bibitem{hanaguri} T. Hanaguri et al, Science {\bf 323}, 923 (2009); 
T. Hanaguri et al, Science {\bf 328}, 474 (2010).

\bibitem{lee} C.-H. Lee et al, J. Phys. Soc. Jpn {\bf 77}, 083704 (2008).

\bibitem{usui} H. Usui and K. Kuroki, Phys. Rev. B {\bf 84}, 024505 (2011).

\bibitem{nagai10} Y. Nagai, K. Kuroki, M. Machida and H. Aoki, 
arXiv:1012.5565.

\bibitem{sakakibarafull} H. Sakakibara, H. Usui, 
K. Kuroki, R. Arita and H. Aoki, Phys. Rev. B {\bf 85}, 064501 (2012).  


\bibitem{bianconi} A. Bianconi et al, J. Phys.: Condens. Matter {\bf 12}, 10655 (2000); 
N. Poccia, A. Ricci and A. Bianconi, 
{\it Advances in Condensed Matter Physics} doi:10.1155/2010/261849 
(2010); N. Poccia et al, nature materials {\bf 10}, 733 (2011). 


\bibitem{rice} T. M. Rice, K.-Y. Yang and F. C. Zhang, 
Rep. Prog. Phys. {\bf 75}, 016502 (2012).

\bibitem{sakai} S. Sakai, Y. Motome and M. Imada, 
Phys. Rev. Lett. {\bf 102}, 056404 (2009); 
Phys. Rev. B {\bf 82}, 134505 (2010).

\bibitem{kubozono_coronene} 
Y. Kubozono et al, 
Phys. Chem. Chem. Phys. {\bf 13}, 16476 (2011).

\bibitem{craciun} M. F. Craciun et al, Phys. Rev. B {\bf 79}, 125116 (2009).

\bibitem{ota} Y. Ota, M. Machida, T. Koyama and H. Aoki, 
Phys. Rev. B {\bf 83}, 060507(R) (2011).  

\bibitem{pandey} S. Pandey, H. Kontani, D. S. Hirashima, 
R. Arita and H. Aoki, arXiv:1107.0122.

\bibitem{tsuji11} N. Tsuji, T. Oka, P. Werner and H. Aoki, 
Phys. Rev. Lett. {\bf 106}, 236401 (2011); arXiv:1110.2925. 

\end{thebibliography}
\end{document}